\documentclass{ws-mpla}
\begin{document}
\markboth{P. T. P. HUTAURUK, ${\it et~al}$.}
{Differential Cross Section Analysis in Kaon Photoproduction}

\catchline{}{}{}{}{}
\title{DIFFERENTIAL CROSS SECTION ANALYSIS IN KAON PHOTOPRODUCTION USING ASSOCIATED LEGENDRE POLYNOMIALS}

\author{\footnotesize P. T. P. HUTAURUK, D. G. IRELAND and G. ROSNER}

\address{Department of Physics and Astronomy, University of Glasgow,\\
Glasgow, G12 8QQ,  Scotland, United Kingdom.\\
p.hutauruk@physics.gla.ac.uk}

\maketitle
\pub{Received (Day Month Year)}{Revised (Day Month Year)}

\begin{abstract}
Angular distributions of differential cross sections from the latest
CLAS data sets ~\cite{bradford}, for the reaction ${\gamma}+p
{\rightarrow} K^{+} + {\Lambda}$ have been analyzed using associated
Legendre polynomials. This analysis is based upon theoretical calculations
in Ref.~\cite{fasano} where all sixteen observables in kaon
photoproduction can be classified into four Legendre classes. Each
observable can be described by an expansion of associated Legendre
polynomial functions. One of the questions to be addressed is how many
associated Legendre polynomials are required to describe the data. In
this preliminary analysis, we used data models with different numbers
of associated Legendre polynomials. We then compared these models by
calculating posterior probabilities of the models.  We found that the
CLAS data set needs no more than four associated Legendre polynomials
to describe the differential cross section data. In addition, we also
show the extracted coefficients of the best model.
\end{abstract}
\ccode{PACS Nos.: include PACS Nos.}

\section{Introduction}
Significant information on the structure of the nucleon can be
obtained by studying its excitation spectrum. Over the last few
decades, a large amount information about the spectrum of the nucleon
has been collected. Most of this information has been extracted from
pion-induced and pion photoproduction reactions. However, pionic
reactions may have biased the information on the existence of certain
resonances. Constituent quark model calculations predict a much richer
resonance spectrum than has been observed in pion production
experiments~\cite{capstick}. Predicted resonances which have not been
observed are called "missing" resonances. Instead, the constituent
quark model also predicts that these "missing" resonances may couple
strongly to  K$\Lambda$ and K$\Sigma$ channels or
other final states involving vector
mesons~\cite{capstick,mart1,mart2}. Since performing kaon-hyperon,
kaon-nucleon or hyperon-nucleon scattering experiments is a daunting
task, kaon photoproduction on the nucleon appears to be a good
alternative solution~\cite{mart1,mart2}.

Experiments on kaon photoproduction and electroproduction started in
the 1960s. However the old experimental data are often inconsistent
and have large error bars. In recent years a large amount of data for
kaon photoproduction has been collected. High statistics
data from CLAS, for differential cross sections, recoil polarization, $C_{x}$ and
$C_{z}$ double polarizations for the reaction $\gamma + p \rightarrow
K^{+} + \Lambda$ have been published~\cite{bradford,bradfor2}.
Additional experimental data have also been measured by
SAPHIR~\cite{glander,tran,glander2}, LEPS~\cite{sumihama,zegers} and
GRAAL~\cite{leres}.

Several previous analyses have been applied to the results of these
experiments, such as
Isobar models~\cite{mart1,mart2,ireland,janssen,janssen2} and 
Coupled channel models~\cite{shyklar,usov,penner}. However different
theoretical model calculations often produce very different
predictions. 

In Ref.\cite{fasano} all sixteen observables in kaon
photoproduction were shown to be classified into the classes 
${\cal L}_0(\hat{{\bf I}};\hat{{\bf E}};\hat{{\bf C_{z'}}};\hat{{\bf
L_{z'}}})$,
${\cal L}_{1a}(\hat{{\bf P}}; \hat{{\bf H}}; \hat{{\bf C_{x'}}}; \hat{{\bf
L_{x'}}})$, 
${\cal L}_{1b}(\hat{{\bf T}}; \hat{{\bf F}}; \hat{{\bf O_{x'}}};
\hat{{\bf T_{z'}}})$ 
and ${\cal L}_2(\hat{{\bf {\Sigma}}}; \hat{{\bf G}}; \hat{{\bf
O_{z'}}}; \hat{{\bf T_{x'}}})$, where each class is an expansion in a
different set of associated Legendre polynomials. What is not apparent
is how many terms in each expansion are required. This work attempts
to address the issue by examining data models with different numbers
of terms, and calculating which one has the greatest posterior
probability. In this article we only focus on the differential cross
section observables, which are described by the associated Legendre
class ${\cal L}_0$.

\section{Analysis Procedure}
\subsection{Data Model}
We construct data models based on Legendre class~${\cal L}_{0}
$. These data models can be written compactly as follows:
\begin{equation}
M^{{\cal L}_{0}}_{L} 
= \sum_{{\it l = 0}}^{{\it l = L}} {\it A_{l}} P_{{\it l0}} (\cos {\theta}). 
\label{eqnmodel}
\end{equation}
where $M^{{\cal L}_{0}}_{L}$ is the data model, and ${\it A_{l}}$ and ${\it
P_{l0}(cos{\theta})}$ are the coefficients
and associated Legendre polynomials. Each data model therefore has a
different ``order'' or maximum number of polynomials. Our task is to
find the most likely order.
\subsection{Model Comparison}
To determine the best model, we evaluate the posterior
probability ~\cite{sivia} for each data model. The ratio of the
probabilities for $M_{L}$ and $M_{0}$ can be written, using Bayes theorem, as follows :
\begin{equation}
R = \frac{P(M_{L}|D)}{P(M_{0}|D)}
= \frac{P(D|M_{L})}{P(D|M_{0})} {\times} \frac {P(M_{L})}{P(M_{0})}. 
\label{eqn2}
\end{equation}
where $P(M_{L}|D)$ is the posterior for the $M_{L}$ model,
$P(D|M_{L})$ is the probability that the data would be obtained,
assuming $M_{L}$ to be true (the likelihood). With no prior
prejudice as to which variant is correct, we obtain the ratio of
likelihoods:
\begin{equation}
R = \frac{P(D|M_{L})}{P(D|M_{0})} \nonumber
\label{eqn3}
\end{equation}
The likelihood $P(D|M_{L})$ is an integral over the joint likelihood
$P(D,\{{\it A_{l}}\}|M_{L})$, where $\{{\it A_{l}}\}$ represents a set
of free parameters:
\begin{eqnarray}
P(D|M_{L}) &=& \int ... \int P(D,\{{\it A_{l}}\}|M_{L}) d^{L}{\it A_{l}}, \nonumber \\
           &=& \int ... \int P(D|\{{\it A_{l}}\},M_{L})P(\{A_{l}\}|M_{L}) d^{L}{\it A_{l}} .
\label{eqn4}
\end{eqnarray}
The function $P({\it A_{l}}|M_{L})$ is the prior probability that the
parameters take on specific values. We assume that each parameter
${\it A_{l}}$ lies in the range ${\it A_{l}^{min}} \leq {\it A_{l}}
\leq {\it A_{l}^{max}}$, and we can write the prior as the reciprocal
of the volume of a hypercube in parameter search space as $P(\{{\it
A_{l}}\}|M_{L}) = \frac{1}{\prod^{L} ({\it A_{l}^{max}} - {\it
A_{l}^{min}})}$ . If the  errors in the data points are Gaussian, it can be shown that
$P(D|\{{\it A_{l}}\},M_{L}) \propto \exp \left( -\frac{\chi^{2}} {2}
\right)$, where  $\chi^{2}$ is the sum of squared residuals. Using a
Taylor series expansion about the minimum $\chi^{2}$ , 
 $\chi^{2}\approx \chi^{2}_{min} + \frac{1}{2} ({\bf X}-{\bf X_{0}})^{T}
{\nabla}^{2}{\chi}^{2} ({\bf X} - {\bf X_{0}}) + ...$, we can write an
approximate form for the likelihood:
\begin{equation}
P(D|M_{L}) \propto \frac{L! (4{\pi})^{L} } {\prod^{L} ({\it A_{l}^{max}} - {\it A_{l}^{min}}) \times \sqrt{Det({\nabla}{\nabla}{\chi}^{2})}}  \exp \left( -\frac{{\chi}^{2}_{min}}{2} \right). 
\label{poste}
\end{equation}
where $L$ is the dimension of the integral and
$(Det({\nabla}{\nabla}{\chi}^{2})$ is the determinant of the Hessian
matrix, which in turn is the inverse of the covariance matrix.

\section{Results}
Using the above analysis procedure, for each of the available photon
energy bins, we fitted each data model to the angular distribution.
This was carried out using the standard minimization package
MINUIT. We then compared models with different numbers of Legendre
polynomials by evaluating Eq.~(\ref{poste}) for each data model.

\begin{figure}[h]
\centerline{\psfig{file=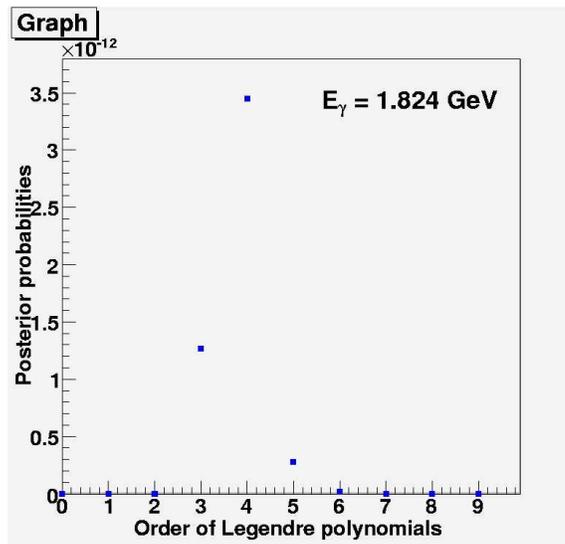,width=3.0in}}
\vspace*{8pt}
\caption{The posterior probabilities for different orders of data
model, for $E_{\gamma}$ = 1.824 GeV.\protect\label{fig1}}
\end{figure}

To illustrate the procedure, we first choose one photon energy
bin at $E_{\gamma}$ = 1.824 GeV as an example. The posterior probabilities are shown
in Fig.~\ref{fig1}, where the order of the data model is shown on the
horizontal axis. The maximum posterior is given by the data model
containing four associated Legendre polynomials. 

\begin{figure}[h]
\centerline{\psfig{file=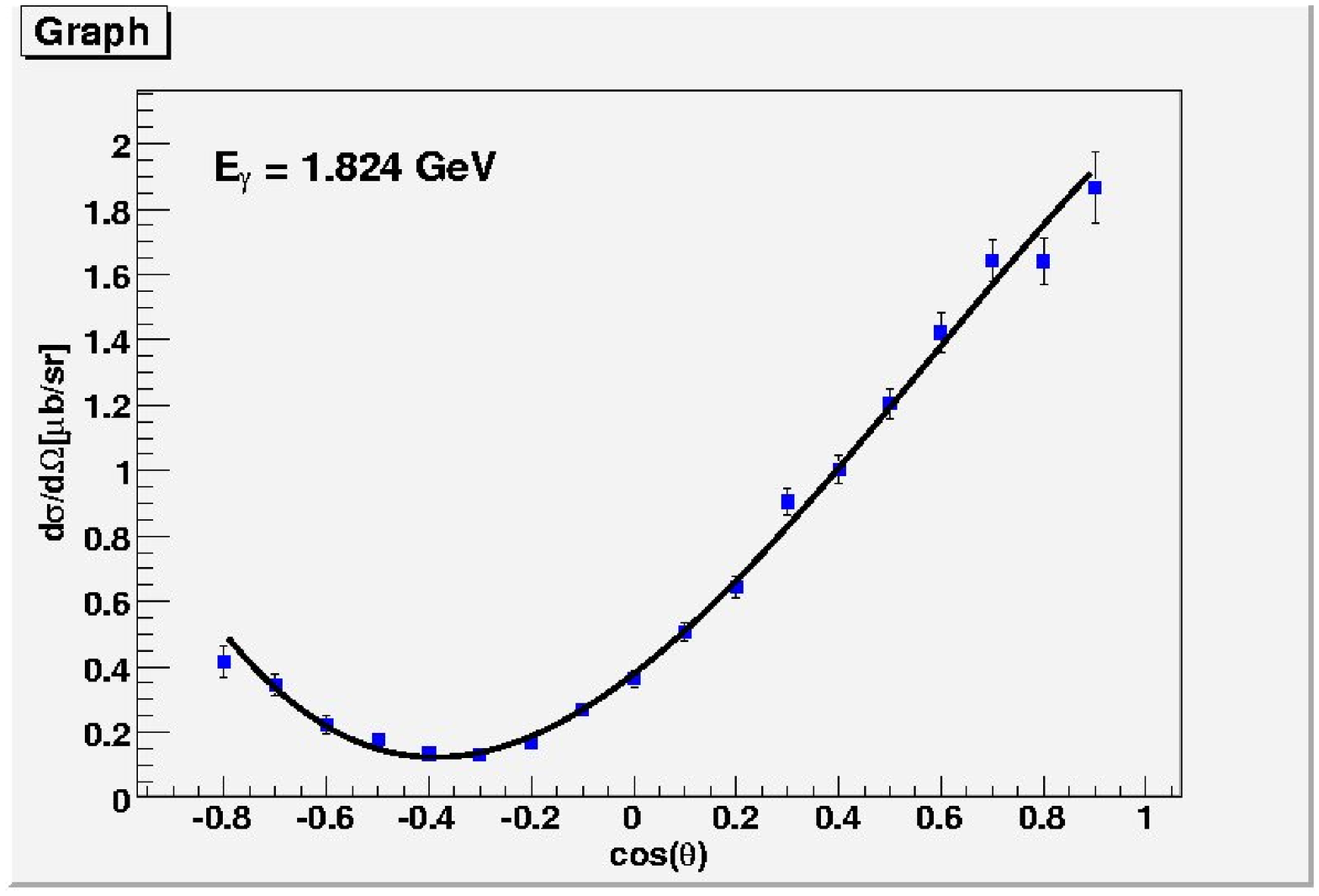,width=3.0in}}
\centerline{\psfig{file=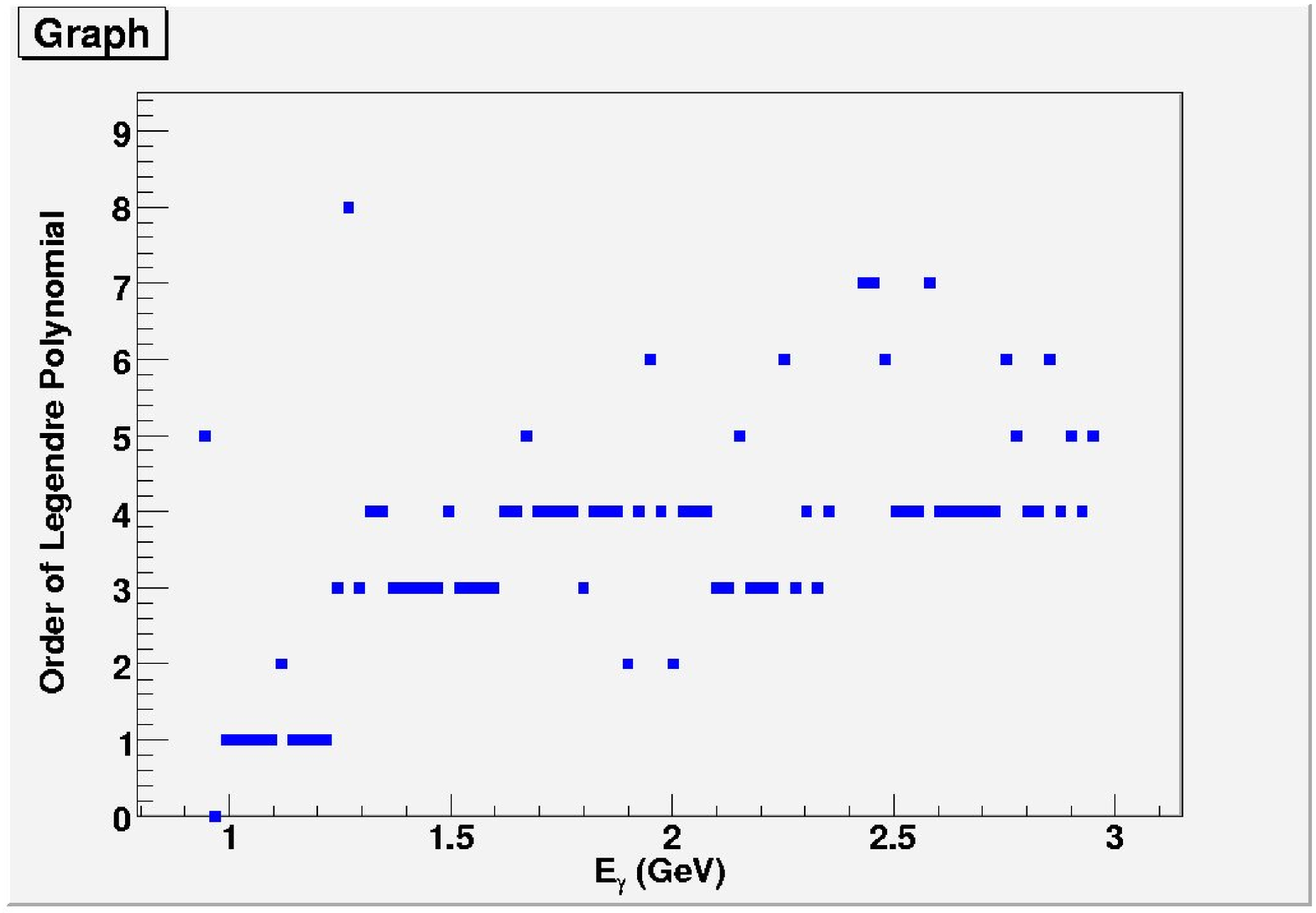,width=3.0in}}
\vspace*{8pt}
\caption{Plot showing the fit of the fourth order data model to the
CLAS cross section data for $E_{\gamma}$ = 1.824 GeV (on the top).The order of associated Legendre polynomial for all photon energy (on the bottom).\protect\label{fig2} }
\end{figure}

\begin{figure}[h]
\centerline{\psfig{file=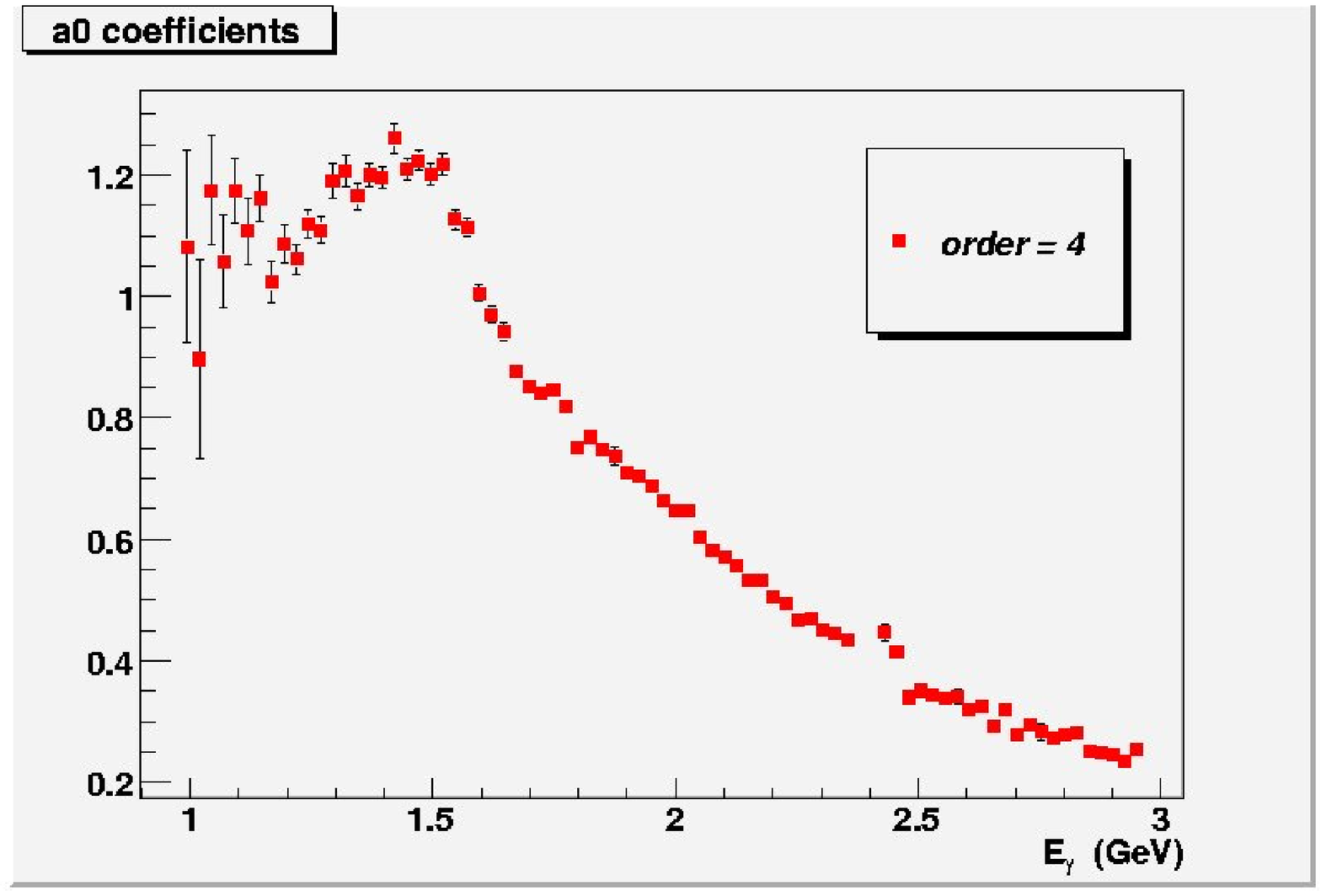,width=2.5in}}
\centerline{\psfig{file=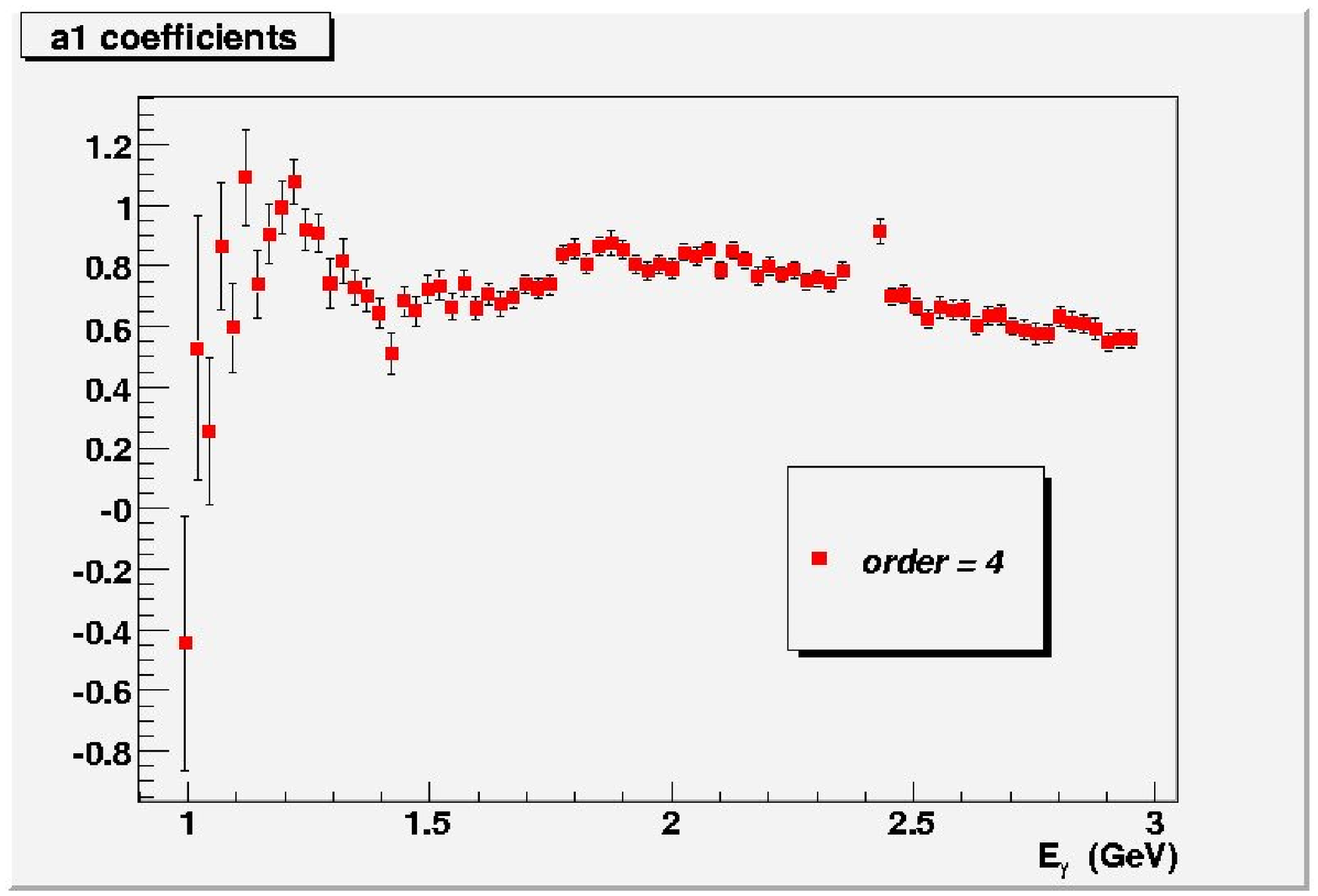,width=2.5in}}
\centerline{\psfig{file=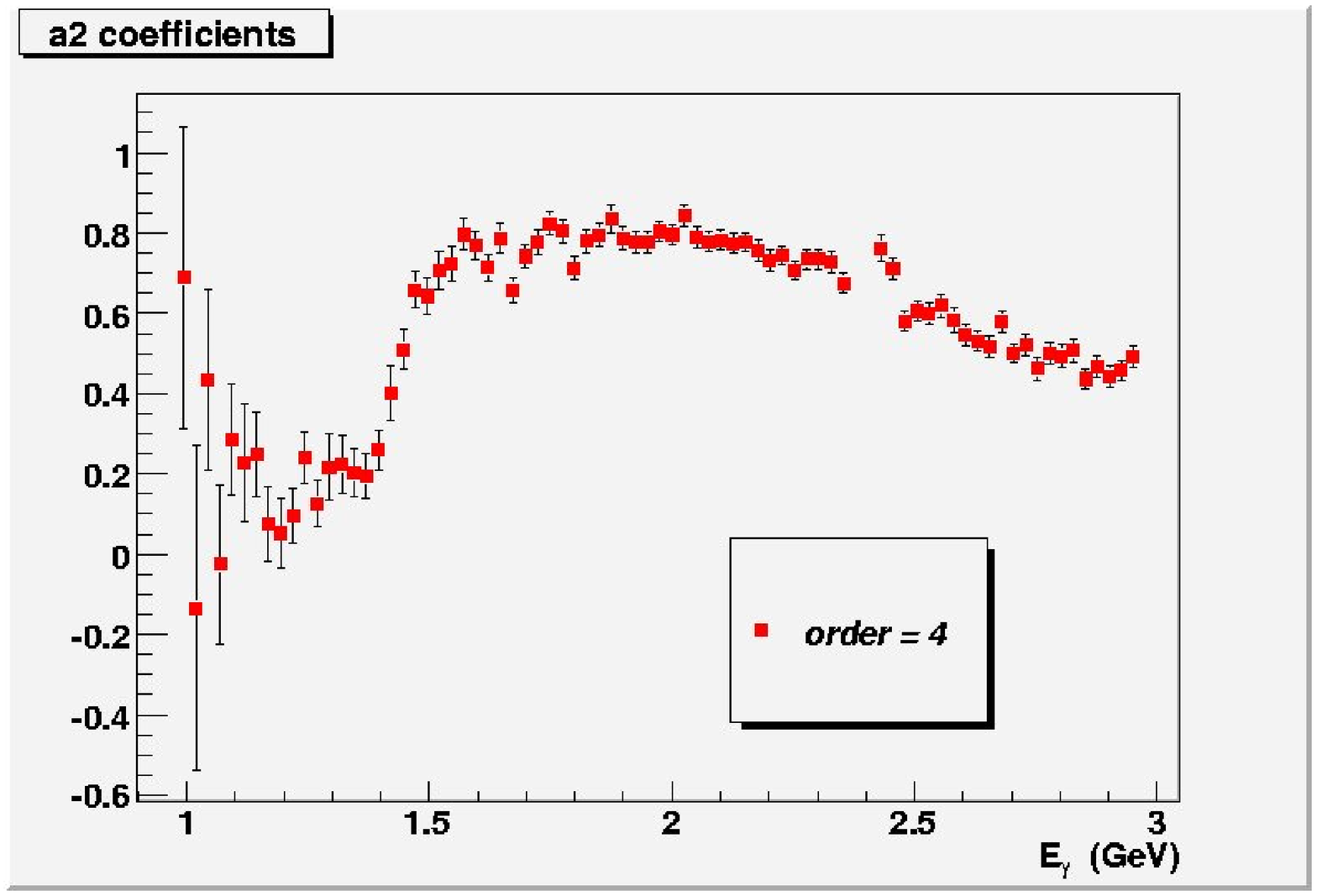,width=2.5in}}
\centerline{\psfig{file=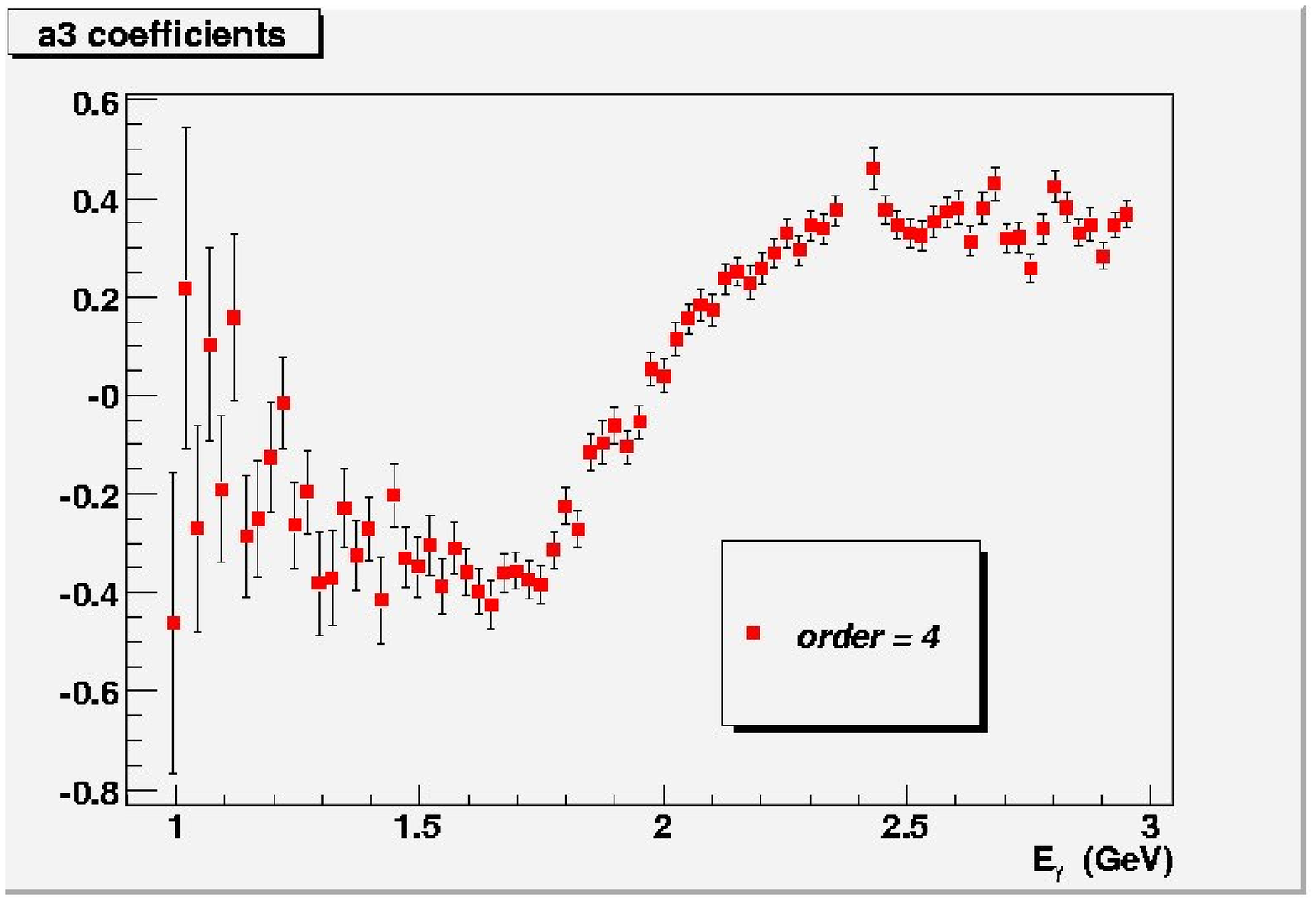,width=2.5in}}
\caption{Extracted associated Legendre polynomial coefficients for each photon energy.\protect\label{fig4}}
\end{figure}

On the left side in Fig.~\ref{fig2} we show the fit of the
fourth order data model to the CLAS differential cross
section data~\cite{bradford} for $E_{\gamma}$ = 1.824 GeV. 
The procedure is repeated for each photon energy bin. The right side in
Fig.~\ref{fig2} the order of data model which has the greatest
probability at each photon energy is plotted. It can be seen that this
generally increase from threshold into the resonance region, but that
the maximum is mostly at the fourth order. 
The distributions of the polynomial coefficients for fourth order data
models as a function of photon energy is shown in Fig.~\ref{fig4}.



\section{Conclusion}
We have analyzed the Legendre polynomial decomposition of differential
cross section data. We generated data models with different numbers of
associated Legendre polynomials. We then compared them by
calculating posterior probabilities. From this analysis,
we found that differential cross section data in this case requires at
least four associated Legendre polynomials.

\section*{Acknowledgments}
This work was supported by SUPA (Scottish Universities Physics Alliance) Fellowship.

\end{document}